\documentclass[twocolumn]{revtex4}
\usepackage{graphicx}
\usepackage{scrtime}
\usepackage{latexsym}
\usepackage{epsfig}
\usepackage{fancyhdr}
\usepackage[hang]{subfigure}
\usepackage{supertabular}
\usepackage{wrapfig}
\usepackage{amssymb}
\usepackage{amsmath}

\begin{document}

%\bibliographystyle{elsart-num}
%\bibliographystyle{iop}
%\begin{frontmatter}

\title{Influence of voltmeter input impedance on quantum Hall effect measurements}

\author{F. Fischer and M. Grayson}
\affiliation{Walter Schottky Institut, Technische Universit\"at M\"unchen, 85748, Garching, Germany}

\begin{abstract}
We report on the influence of voltmeters on measurements of the longitudinal resistance in the quantum Hall effect regime. 
We show that for typical input resistances for standard digital lock-in amplifiers the longitudinal resistance can show a non-zero minimum which might be mistaken for parallel conduction in the doping layer. This residual
resistance can be calculated with $R^{res}_{xx}=R_{xy}^2/R_{in}+j\omega C R_{xy}^2$, where $R_{in}$ is the input resistance of the voltmeter and C the measurement capacitance. 
%a too low input resistance of the voltmeter affects the longitudinal resistance
%like it would be affected by a parallel conducting layer, but 
In contrast to a real parallel conduction the effect disappears when either the current source and ground contact are swapped or the polarity of the B-field is changed. 
%when the polarity of either the magnetic field or the current is changed. 
We discuss the influence of input capacitances and stray capacitances on the measurement. 
%From measurements of the longitudinal resistance we are able to deduce the input impedance of a lock-in. 
%We give an overview over the input impedances of commercially available voltmeters.
The data demonstrates the influence of the voltmeter input impedance on the longitudinal resistance measurement.
\end{abstract}

%\keywords
%\pacs{}

\maketitle

\section{Introduction}
The quantum Hall effect (QHE) \cite{klitzing:494,klitzing:519} of a two-dimensional system (2DS) in a perpendicular magnetic field B is characterized by the quantization of the transversal resistance to $R_{xy}=h/\nu e^2$ ($\nu=1,2,...$) coincident with minima in the longitudinal resistance $R_{xx}$ in the quantum Hall regime. $R_{xx}$ vanishes to $R_{xx}$=0 in high mobility samples and sufficiently low temperatures. %(see Fig. \ref{rxx_rxy}). 
A similar behaviour occurs for the fractional quantum Hall effect (FQHE) \cite{stormer:1953,stormer:S298} for odd denominator fractions ($\nu = \frac{1}{3}$, $\frac{2}{5}$, $\frac{2}{3}$, ...). 
%While a non-vanishing $R_{xx}$ can also be due to a low mobility $\mu$ or a too high temperature $T$, at a low enough temperature T and high mobility $\mu$ it 
A flat low temperature non-zero minimum is normally attributed to a parallel conducting dopant layer \cite{grayson:1,burgt:12218,contreras:1251}.\\
% This is especially evident if the width $\Delta B$ of a $R_{xx}$ minimum is $\Delta B>0$ while $R_{xx}>0$ in this intervall.\\
%\begin{figure}[!h]
%\center
%\includegraphics[width=7cm,keepaspectratio]{graphics/rxxrxy.eps}
%\caption{Hall resistance $R_{xy}$ and longitudinal resistance $R_{xx}$ measured in a two-dimensional eletron gas (2DEG) in a AlGaAs/GaAs heterostructure with a mobility of $\mu=$ at a density of $n=$ (temperature in den graphen).}
%\label{rxx_rxy}
%\end{figure}
\indent In the following article we will show measurements where parallel conduction seems to be evident but in reality is due to a measurement error, because of a finite input impedance of the measuring voltmeter.
\section{Analysis}
\begin{figure}[!h]
\center
\includegraphics[width=6cm,keepaspectratio]{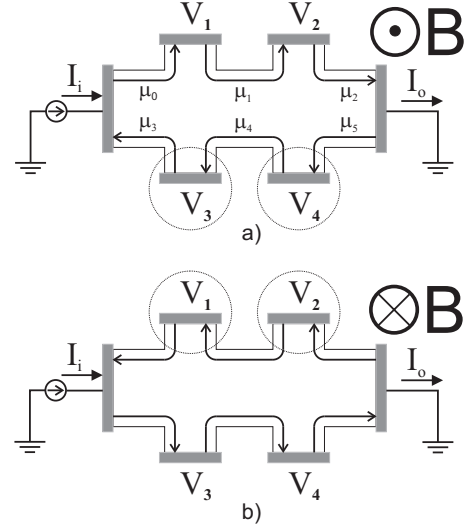}
\caption{Typical example of a Hall-bar geometry. $I_i$ and $I_o$ indicate the input and output currents. Figure a: For the indicated B-field pointing out of the plane the arrow-lines represent electrons in the QHE regime travelling clockwise along the edge having a certain chemical potential $\mu_i$ ($i=1,..,6$). For hole edge channels one can simply flip the indicated B-field polarity. Figure b: For the B-field pointing into the plane electrons travel counterclockwise.\\ The circled contacts show proper $R_{xx}=0$ minima, whereas the voltage contacts opposite show residual $R_{xx}^{res}$ in the minima described in Eqs. (\ref{real_resistance}) and (\ref{im_resistance})}
\label{hall_bar}
\end{figure}

\begin{figure}[!h]
\center
\includegraphics[width=5cm,keepaspectratio]{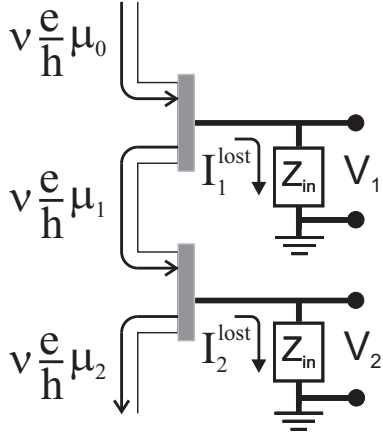}
\caption{Circuit diagram for a measurement of the longitudinal resitance. A voltmeter with input impedance $Z_{in}$ measures the differential voltage $V_{xx}=V_1-V_2$.}
\label{input_impedance}
\end{figure}

The standard method to measure the quantum Hall effect is a four-point measurement with a Hall-bar indicated in Fig. \ref{hall_bar}. A typical Hall-bar has a current input $I_i$ and a current output $I_o$. There are 4 voltage contacts $V_1$, $V_2$, $V_3$ and $V_4$ to measure the electrical potential at fixed points along the sample. For an ideally homogeneous and symmetric sample with $I_i=I_o=I$, $R_{xy}$ is given by $R_{xy}=|(V_1-V_3)|/I=|(V_2-V_4)|/I$ and $R_{xx}=|(V_1-V_2)|/I=|(V_3-V_4)|/I$.  
In the common QHE edge state picture \cite{halperin:2185} the current flows only along a small strip at the edge of the sample as indicated in Fig. (\ref{hall_bar}) for electrons. Following
the Landauer-B\"uttiker formalism for QHE edge channels \cite{jain:4276,buttiker:9375} the total current along the edge is given by 
\begin{equation}
\nu \frac{e}{h}\mu_k = \nu \frac{e^2}{h}V_k
\label{current_potential_relation}
\end{equation}
with $\nu$ being the filling factor, $\mu_k$ the chemical potential of the edge channel with index k and $V_k$ the corresponding voltage. The chemical potential of each edge channel is determined by the voltage of the injecting contact. 
In contrast to the conventional current, the direction of the edge channel flow equals the direction of the carrier movement and is determined by the sign of the charge and the polarity of the magnetic field. The following discussion assumes edge currents running clockwise, as in the case of electrons and a magnetic field pointing out of the Hall-bar (see Fig. \ref{hall_bar}).\\ 
\indent If a current is driven through the Hall-bar but the voltage contacts are not connected then $\mu_0=\mu_1=\mu_2$ is the chemical potential of the current source and $\mu_3=\mu_4=\mu_5$ is the chemical potential of the ground. The same is applicable if the measurement devices connected to the voltage contacts $V_i$ do not draw any current, and the edge current flowing into a voltage contact equals the edge current flowing out.
In the QHE regime $R_{xx}$ is then given by
\begin{subequations}
\begin{eqnarray}
R^{1-2}_{xx}&=&\frac{V_1-V_2}{I}=e\frac{\mu_1-\mu_2}{I}=0~~\textnormal{and}\\
R^{3-4}_{xx}&=&\frac{V_3-V_4}{I}=e\frac{\mu_3-\mu_4}{I}=0~.
\end{eqnarray}
\end{subequations}
\\
When the voltage contacts 1 and 2 are connected to a voltmeter with a finite input-impedance $Z_{in}= (R_{in}^{-1}+j\omega C)^{-1}$ a current $I^{lost}$ flows through the voltmeter to ground (see Fig. \ref{input_impedance}).
Since the potential of a contact is given by the chemical potential of the reduced outflowing edge current, this leads to a change of the chemical potential of the edge current at every voltage contact. One would then measure a finite residual resistance $R_{xx}^{res}\not= 0$ in a QHE minimum of $R_{xx}$.
The measured voltages $V^{1-2}_{xx}$ and $V^{3-4}_{xx}$ are given by
\begin{subequations}
\begin{eqnarray}
V_{xx}^{1-2}&=&e(\mu_1-\mu_2)=\frac{h}{\nu e^2}I^{lost}_2=\frac{h}{\nu e^2}\frac{1}{Z_{in}}e\mu_2 \\
V_{xx}^{3-4}&=&\frac{h}{\nu e^2}\frac{1}{Z_{in}}e\mu_3
\end{eqnarray}
\end{subequations}
with $(\nu e/h) \mu_1 = (\nu e/h) \mu_2 + I^{lost}_2$ required by current conservation and $I^{lost}_2=V_2/Z_{in}$.\\
\indent Measurements of the QHE are normally performed with one of the current contacts set to ground and driving a constant DC or constant AC sinusoidal current through the sample. In the QHE regime the voltage of the non-grounded current contact is given by $V_{xy}=R_{xy}I=h/\nu e^2I$ with I being the driving current.
By eliminating $\mu_1$ in the set of equations:
\begin{subequations}
\begin{eqnarray}
e\mu_0 &=& I_1^{lost} R_{xy}+e\mu_1 \label{knots1}\\
e\mu_1 &=& I_2^{lost} R_{xy}+e\mu_2
\end{eqnarray}
\end{subequations}
one deduces
\begin{eqnarray}
\mu_2=\mu_0 \left(\frac{R_{xy}^2}{Z_{in}^2}+2\frac{R_{xy}}{Z_{in}}+1\right)^{-1} \\
\end{eqnarray}
respectively
\begin{eqnarray}
\mu_3=\mu_5 \left(\frac{R_{xy}^2}{Z_{in}^2}+2\frac{R_{xy}}{Z_{in}}+1\right)^{-1}
\end{eqnarray}
%For $I_{lost}<< I$ is $V_{xy}\approx e\mu_0\approx e\mu_1\approx e\mu_2$ leading to
In standard voltmeters $R_{xy} \ll Z_{in}$, since $R_{xy} \sim h/e^2=25~\textnormal{k}\Omega$, and $Z_{in} \ge 10~\textnormal{M}\Omega$. The above expression reduces to $\mu_2 \approx \mu_0=V_{xy}/e$ and $\mu_3 = \mu_5=0$, thus
\begin{subequations}
\begin{eqnarray}
\label{vxx1}
V_{xx}^{1-2}=R_{xy}^2 \frac{1}{Z_{in}} I&=&\left(\frac{h}{\nu e^2}\right)^2 \frac{1}{Z_{in}} I \nonumber \\
&=& R_{xy}^2 (\frac{1}{R_{in}}+j\omega C)I \label{result1}\\
V_{xx}^{3-4}&=&0 \label{result2}
\end{eqnarray}
\end{subequations}
the real and imaginary components of $V_{xx}^{1-2}$ are then given by
\begin{subequations}
\begin{eqnarray}
Re\left(V_{xx}^{1-2}\right)&=& \frac{R_{xy}^2}{R_{in}}I \nonumber \\
\Rightarrow
Re\left(R_{xx}^{1-2}\right)&=& \frac{Re(V_{xx})}{I}=\frac{R_{xy}^2}{R_{in}}  \label{real_resistance}\\
Im\left(V_{xx}^{1-2}\right)&=&R_{xy}^2\omega C I \nonumber \\
\Rightarrow
Im\left(R_{xx}^{1-2}\right)&=&\frac{Im(V_{xx})}{I}=R_{xy}^2\omega C 
\label{im_resistance}
\end{eqnarray}
\end{subequations}
%For $R=10~\textnormal{M}\Omega$ is $\textnormal{Re}\left(R_{xx}\right)= (1/\nu)66.6~\Omega$. For the $R_{xx}$ oscillations shown in Fig. \ref{rxx_rxy} this would be a reasonable contribution, destroying the validity of the measurement. 
%The voltage contacts which are approximatly on ground potential don't show this behaviour. 
\indent The results (\ref{result1}) and (\ref{result2}) show that on one side of the Hall-bar there is a spurious measured resistance and on the other side there is none. Whether the residual resistance $R^{res}_{xx}=R_{xy}^2/R_{in}+j\omega C R_{xy}^2$ is measured is determined by 
the chemical potential sourcing the edge currents.
%the direction of the edge currents, which depend only upon the sign of the charge and the polarity of the magnetic field. 
If an effect of the input impedance on the measurement is suspected, one can investigate that by either swapping the current source contact with the ground contact, by changing the polarity of the magnetic field or  by using voltage contacts on the other side of sample.
It is worth reminding the reader that a parallel conducting layer would also lead to a non-zero $R_{xx}$ minima increasing with the magnetic field \cite{grayson:1}, but {\it would be seen at \underline{all} $R_{xx}$ contact pairs, would not disappear upon reversing the B-field and is normally not quadratic in $1/\nu$}.\\
%The influence of the input impedance on measurements of the transversal resistance $R_{xy}$ is much weaker. With equation (\ref{knots1}) one derives for the measured transversal resistance $R_{xy}^{meas}$
%\begin{equation}
 
\section{Measurement}
A voltage measurement can either be done DC or AC, the latter normally performed with lock-in amplifiers. Table \ref{voltmeters} gives
\begin{table*}
\center
\begin{tabular}{|c|c|c|c|c|c|}
\hline
   manufacturer & model & mode & R & C & residual Re($R_{xx}$) \\
\hline
Keithley & 1801           & DC           & $>1$ G$\Omega$ & - & $<0.66~\Omega/\nu^2$\\
\hline
Agilent & 34420A           & DC & $>10$ G$\Omega$          & $<3.6$ nF & $<0.06~\Omega/\nu^2$\\
\hline
Signal Recovery (EG\&G) & 7260           & digital AC & $10$ M$\Omega$          & $30$ pF & $66.6~\Omega/\nu^2$\\
\hline
Signal Recovery (EG\&G) & 210           & analog AC & $100$ M$\Omega$          & $25$ pF & $6.66~\Omega/\nu^2$\\
\hline
Stanford Research & 810           & digital AC       & $10$ M$\Omega$          & $25$ pF & $66.6~\Omega/\nu^2$\\
\hline
Stanford Research & 510           & analog AC & $100$ M$\Omega$          & $25$ pF & $6.66~\Omega/\nu^2$\\
\hline
\end{tabular}
\label{voltmeters}
\caption{Input impedances of different commercially available voltmeters (data taken from the product specifications).}
\end{table*}
 an overview over the input impedances $Z_{in}= (R_{in}^{-1}+j\omega C)^{-1}$ of commercially available voltmeters. 
 %We performed our measurements with an EG\&G 7260 and a Standford SR810 lock-in amplifier. 
\begin{figure}[!h]
\center
\includegraphics[width=7cm,keepaspectratio]{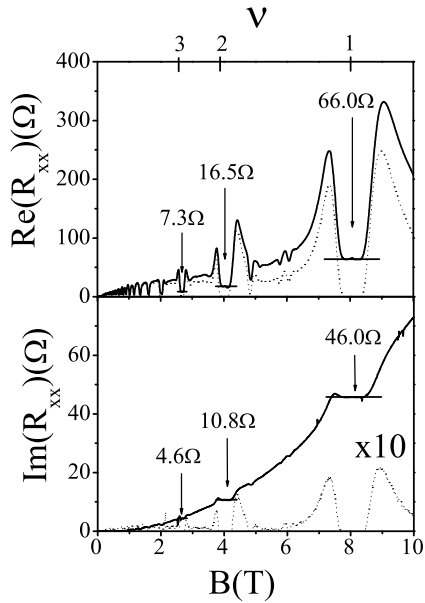}
\label{rxx_real_and_im}
\caption{Real part (top figure) and imaginary part (bottom figure) of $R_{xx}$ 
corresponding to $R_{xx}^{1-2}$ (solid line) and $R_{xx}^{3-4}$ (dotted line) as measured in Fig. \ref{hall_bar}a)}
%with positive (dotted line) and negative (solid line) polarity of the current.}
\end{figure}
%\begin{figure}[!t]
%\center
%\includegraphics[width=7cm,keepaspectratio]{graphics/parallel_channel_analysis_im.EPS}
%\caption{Imaginary part of $R_{xx}$ with positive (dotted line) and negative (solid line) polarity of the current.}
%\label{rxx_im}
%\end{figure}
The data in Fig. \ref{rxx_real_and_im} shows the real and imaginary parts of
$R_{xx}^{1-2}$ (solid line) and $R_{xx}^{3-4}$ (dotted line) for a 2DEG in an AlGaAs/GaAs heterostructure measured as in Fig. \ref{hall_bar}a) with B coming out of the sample. The data was taken with an EG\&G 7260 lock-in amplifier in a van-der Pauw geometry instead of a Hall-Bar geometry at an excitation frequency of $\omega = 2 \pi f= 17~\textnormal{Hz}$. 
From the finite $R_{xx}^{1-2}$ resistance minima one can directly deduce the input impedance of the lock-in amplifier using eq. (\ref{result1}) $Z_{in}=10.1~\textnormal{M}\Omega+j\omega4~\textnormal{nF}$. While the real part is nearly exactly the value given in Table \ref{voltmeters}, the capacitance from the imaginary part is about a factor of $\times20$ larger than the lock-in specifications. This is because the total capacitance $C_{total}=4~\textnormal{nF}$ is $C_{total}= C_{lock-in}+C_{setup}$, where $C_{lock-in}=30~\textnormal{pF}$ is the capacitance of the lock-in given in Table \ref{voltmeters} and $C_{setup}=3970~\textnormal{pF}$ the stray capacitance of the rest of the measurement setup principally the probe wiring. Since $\textnormal{Im}(R_{xx}) \sim \omega$ scales linearly with frequency, any structures in $R_{xx}$ that change frequency can be assumed to derive from the total measurement capacitance.\\
\indent Standard lock-in amplifiers are equipped with optional internal line filters at frequencies 50 Hz and 100Hz (or 60 Hz and 120 Hz depending on the line standard). 
%additional series impedances into the measurement line. This can lead to 
If activated, these filters can even at low lock-in frequencies (1-10 Hz), induce
a phase shift of the residual resistance signal $R_{xx}^{res}=R_{xx}^0e^{i\phi}$ by $\Delta\phi$ to  
$R_{xx}=R_{xx}^0e^{i\left(\phi+\Delta\phi\right)}$ and rotating some of the imaginary part into the real. Therefore a large capacitive signal can be rotated into the real $R_{xx}$ trace and vice versa. Because this can lead to a deviation of the expected residual resistance values
it is recommended not to use internal filters unless they are needed, and then to carefully calibrate out the induced phase shift.
%But there are a couple of tests which can rule out the input impedance as the reason for the parallel conduction. \\

\section{Conclusion}
We conclude that special care has to be taken while setting up a measurement of the longitudinal resistance.
We showed that although the value of the voltmeter input resistance  of $R=10~\textnormal{M}\Omega$ seems to be sufficiently high, it can highly influence a measurement of $R_{xx}$ by showing a residual longitudinal resistance of $R^{res}_{xx}=R_{xy}^2/R_{in}+j\omega C R_{xy}^2$ and lead to the false assumption of a parallel channel. If a finite $R_{xx}$ minimum drops to zero with any of the following 3 tests, the input resistance is responsible for the residual $R_{xx}$. 
\begin{enumerate}
\item Switching the polarity of the magnetic field.
\item Swapping the current source contact.
\item Using voltage contacts on the other side of the sample.
\item For phasing errors, one can check if changing the excitation frequency by a certain factor, changes the $R_{xx}$ signal at the minimum. 
\end{enumerate}
\indent \indent In contrast to AC lock-in amplifiers, commercially available DC-voltmeters usually have sufficiently high input impedances of $R > 1~\textnormal{G}\Omega$ giving a residual $R_{xx}^{res} = 0.66~\Omega/\nu^2$ (see Table \ref{voltmeters}), a value small enough that it does not disturb a measurement. The analog lock-in amplifiers have a residual $Re(R_{xx}^{res}) = 6.6~\Omega/\nu^2$, a value which could already be seen in measurements especially in the FQHE-regime. The digital lock-in amplifiers have residual input resistances of $Re(R_{xx}^{res})
 = 66.6~\Omega/\nu^2$, a value which can clearly be seen in measurements. It is advisable to make use of pre-amplifiers with higher input impedances.  
%Although the input capacitance of the lock-in amplifiers with $C < 30~\textnormal{pF}$ at a residual $Im(R_{xx}) < 0.2~\Omega$ at 10 Hz excitation frequency is reasonable low, the capacitance $C_{setup}$ is $C_{setup} >> C_{lock-in}$ for typical measurement setups. 
The total capacitance of the measurement setup $C_{setup}$ (typically several hundred pF or more) must be considered when checking for $Im(R_{xx}^{res})$ effects.
Built-in line filters in dilution refrigerator systems to filter out radio-frequencies have a larger contribution to $C_{setup}$.\\
\indent In summary, the ideal setup has a very high input impedance preamplifier to the lock-in with a low input capacitance and small stray capacitances. As a rule of thumb, one should {\it always} know the chirality of the edge states and measure $R_{xx}$ at the ground-potential side of the sample.
 
\begin{acknowledgements}
This work was supported financially by Deutsche Forschungsgemeinschaft via Schwerpunktprogramm Quantum-Hall-Systeme. M. Grayson would like to thank A. v. Humboldt Foundation for support.
\end{acknowledgements}

\end{document}